\documentclass[aps,prl,twocolumn,floatfix]{revtex4}

\usepackage{graphics,graphicx}
\usepackage{bm,bbm}
\usepackage{amssymb,amsmath,amsfonts}
\usepackage{mathrsfs}
\usepackage{calc}

\def\jpa#1#2#3{J.~Phys.~A,~{\bf #1},\ #2,\ (#3)}

\def\prl#1#2#3{Phys.~Rev.~Lett.,~{\bf #1},\ #2,\ (#3)}

\def\pla#1#2#3{Phys.~Lett.~A,~{\bf #1},\ #2,\ (#3)}

\def\etal{{\it et al.}\,}

\def\bpmat{\begin{pmatrix}}
\def\epmat{\end{pmatrix}}
\def\bmat{\begin{matrix}}
\def\emat{\end{matrix}}

\def\1{\mbox{1\hskip-.25em l}}
\def\PT{$\mathcal{PT}\,$}
\def\beq{\begin{equation}}
\def\eeq{\end{equation}}
\def\beqar{\begin{eqnarray}}
\def\eeqar{\end{eqnarray}}

\begin{document}

\title{\bf Visualization of Branch Points in $\mathcal{PT}$-Symmetric Waveguides}
\author{Shachar Klaiman}
\affiliation{Shulich Department of Chemistry and Minerva Center
for Nonlinear Physics of Complex Systems, Technion -- Israel
Institute of Technology, Haifa 32000, Israel.}
\author{Nimrod Moiseyev}
\affiliation{Shulich Department of Chemistry and Minerva Center
for Nonlinear Physics of Complex Systems, Technion -- Israel
Institute of Technology, Haifa 32000, Israel.}
\author{Uwe G\"{u}nther}
\affiliation{Research Center Dresden-Rossendorf, POB 510119,
D-01314 Dresden, Germany}
\begin{abstract}
The visualization of an exceptional point in a \PT symmetric
directional coupler(DC) is demonstrated. In such a system the
exceptional point can be probed by varying only a single
parameter. Using the Rayleigh-Schr\"{o}dinger perturbation theory
we prove that the spectrum of a \PT symmetric Hamiltonian is real
as long as the radius of convergence has not been reached. We also
show how one can use a \PT symmetric DC to measure the radius of
convergence for non \PT symmetric structures. For such systems the
physical meaning of the rather mathematical term: radius of
convergence, is exemplified.
\end{abstract}
 \maketitle

In the past several years, following the seminal paper by Bender
and Boettcher \cite{benderprl1}, non-hermitian \PT-symmetric
Hamiltonians have caught a lot of attention (see \cite{benderMKS}
and references therein). Under certain conditions \PT - symmetric
Hamiltonians can have a completely real spectrum and thus can
serve, under the appropriate inner products, as the Hamiltonians
for unitary quantum systems \cite{mostafazadehprl}.

Recently, the realization of \PT - symmetric "Hamiltonians" has
been studied using optical waveguides with complex refractive
indices \cite{Muga,Christo}. The equivalence of the Maxwell and
Schr\"{o}dinger equations in certain regimes provides a physical
system in which the properties of \PT-symmetric operators can be
studied and exemplified.

An extremely interesting property of \PT - symmetric operators
stems from the anti-linearity of the time symmetry operator.
Consider a \PT - symmetric operator $\hat{H}$, i.e.,
$[\mathcal{PT} , \hat{H}]=0$. Due to the non-linearity of
$\mathcal{T}$ one \textit{cannot} in general choose simultaneous
eigenfunctions of the operators \PT and $\hat{H}$. However, if an
eigenvalue of the $\hat{H}$ is real then it's corresponding
eigenfunction is a also an eigenfunction of the \PT operator. This
property has come to be known as exact/spontaneously-broken  \PT -
symmetry. Exact \PT - symmetry refers to the case when every
eigenfunction of the \PT symmetric operator is also an
eigenfunction of the \PT operator. In any other case the \PT -
symmetry is said to be broken.

Usually, the transition between exact and spontaneously-broken \PT
symmetry can be controlled by a parameter in the Hamiltonian. This
parameter serves as a measure of the non-hermiticity. An important
class of \PT - symmetric Hamiltonians are of the form:
$\hat{H}(\lambda)=H_0+i\lambda V$. Where $H_0$(and $V$) are real
and symmetric(anti-symmetric) with respect to parity so that
$[\mathcal{PT} , \hat{H}]=0$. When $\lambda=0$ the Hamiltonian is
hermitian and the entire spectrum is real. The spectrum remains
real even when $\lambda\neq0$ as long as $\lambda<\lambda_c$. At
this critical value and beyond, pairs of eigenvalues collide and
become complex, see for example \cite{Znojil}. Bender \etal
\cite{BBM} showed that the reality of the spectrum is explained by
the real secular equations one can write for \PT - symmetric
matrices. These secular equations will depend on the
non-hermiticity parameter and, consequently, yield either real or
complex solutions. Delabaere \etal \cite{Delabaere} showed for the
one-parameter family of complex cubic oscillators that pairs of
eigenvalues cross each other at Bender and Wu branch points. Dorey
\etal \cite{Doreyproof} after proving the reality of the spectrum
for a family of \PT-symmetric Hamiltonians, showed \cite{Dorey}
that at the point where the energy levels cross, i.e., the
critical value of $\lambda$, a super-symmetry is broken and not
only the eigenvalues but also the eigenfunctions become the same.

Consider the family of Hamiltonians given by: $\hat{H}=H_0+\lambda
V=H_0+i|\lambda| V $. The existence of a branch point in the
Hamiltonian's spectrum determines the radius of convergence of a
series expansion of the energy in $\lambda$. Friedland and one of
us \cite{NMF} proved that for two real symmetric matrices
$\bf{H_0}$ and $\bf{V}$ that do not commute there exists
\textit{at least} one branch point $\lambda_{bp}$ for which
$\frac{dE}{d\lambda}|_{\lambda=\lambda_{bp}}=\infty$. Therefore,
the expansion of the energy in powers of $\lambda$ converges only
as long as $|\lambda|<|\lambda_{bp}|$. The most common situation
is when the branch point is associated with the coalescence of two
eigenfunctions and the two corresponding eigenvalues. Such a point
in the spectrum is often referred to as an \textit{exceptional
point} \cite{Kato}. Exceptional points in physical systems have
been studied, e.g., in
\cite{Berry,Cartarius,Rubinstein,Uwe,Cejnar}. Recently,
exceptional points have been observed experimentally in microwave
cavities \cite{Dembowski}. In general, as stated in the theorem
above, the value of the parameter $\lambda$ at which the branch
point occurs is a complex number. This demands the control of the
Hamiltonian by at least two parameters \cite{Cartarius}. From the
evidence of branch points in previous studies of \PT-symmetric
Hamiltonians, e.g., \cite{Znojil}, it is plausible to assume that
the branch point is located on the imaginary axis, i.e.,
$\lambda_{bp}=i|\lambda_{bp}|$. Hence, the two parameter
dependence reduces to a dependence on a single real parameter.
Therefore, the study of exceptional points in \PT - symmetric
systems is strongly simplified.

We restrict the discussion to the cases for which
$|\lambda_{bp}|\neq0$. Thus for $\lambda=i|\lambda|$ and
$\lambda_{bp}=i|\lambda_{bp}|$ in the vicinity of an exceptional
point one can write,
\begin{eqnarray}
E(\lambda)&=& E^{bp}\pm D\sqrt{(\lambda-\lambda_{bp})(\lambda-\lambda_{bp}^*)}\\
&=& E^{bp} \pm D\sqrt{|\lambda_{bp}|^2-|\lambda|^2}\,.
\end{eqnarray}
If $|\lambda| > |\lambda_{bp}|$ one gets a pair of complex
conjugate eigenvalues $E$ and $E^*$, whereas in the case for which
$|\lambda| < |\lambda_{bp}|$ the system has two real eigenvalues.
Therefore, $\lambda_{bp}$ corresponds to the critical value,
$\lambda_c$, at which the transition between exact and
spontaneously broken \PT symmetry occurs. An interesting
consequence of the above analysis is that if one treats the
potential $\lambda V$ as a perturbation the
Rayleigh-Schr\"{o}dinger perturbation expansion would converge as
long as $|\lambda| < |\lambda_{bp}|$. One would then expect that
the reality of the spectrum should be apparent from such an
expansion. We shall prove this in the following.

Consider the following time-independent Schr\"{o}dinger equation:
\begin{equation}
(H_0+\lambda V)\Psi_j(\lambda)=E_j(\lambda)\Psi_j(\lambda),
\end{equation}
Where again, $H_0$(and $V$) are real and symmetric(anti-symmetric)
with respect to parity and are both hermitian operators. Assuming
that $|\lambda| < |\lambda_{bp}|$, one can expand the eigenvalues
and eigenfunctions in a \textit{convergent} power series in powers
of $\lambda$:
\begin{eqnarray}
E_j(|\lambda| < |\lambda_{bp}|)&=&\sum_{n=0}^\infty \lambda^nE_j^{(n)}, \\
\Psi_j(|\lambda| < |\lambda_{bp}|)&=&\sum_{n=0}^\infty \lambda^n \Psi_j^{(n)},
\end{eqnarray}
where $E_j^{(n)}\,(\Psi_j^{(n)})\,;\,n=0,1,2..$ are the {\it real}
energy (wavefunction) correction terms of the
Rayleigh-Schr\"{o}dinger perturbation expansion. The (2n+1)- rule
stated by  Wigner \cite{wigner} implies that,
\begin{equation}
E_j = \langle\chi_j^{(n)}|H_0+\lambda V| \chi_j^{(n)}\rangle
+O(\lambda^{2n+2})
\end{equation}
where,
\begin{equation}
\chi_j^{(n)}(x)=\sum_{k=0}^n \lambda^k \psi_j^{(k)}
\end{equation}
and $\langle\chi_j^{(n)}|\chi_j^{(n)}\rangle=1$. Therefore,
following the (2n+1)-rule,
\begin{equation}
E_j^{(2n+1)}=\langle\psi_j^{(n)}|V|\psi_j^{(n)}\rangle
\label{E-odd}
\end{equation}
where the n-th order correction to the exact eigenfunction
$\Psi_j$ is the solution of the following equation,
\beqar
\psi_j^{(n)}(x)&=&
G_0'(E_j^{(0)})V(x)|\psi_j^{(n-1)}\rangle\\
\nonumber &-&\sum_{k=1}^{n-1}E_j^{(k)}G_0'(E_j^{(0)})
|\psi_j^{(n-k)}\rangle,
\label{psi-n}
\eeqar
and $G_0'(E_j^{(0)})=\sum_{q\ne j}\langle
x|\psi_q^{(0)}\rangle\langle \psi_q^{(0)} | (
E_j^{(0)}-E_q^{(0)})^{-1}$. The parity symmetry of the
zeroth-order Hamiltonian ensures that its eigenfunction
$\psi_q^{(0)}(x)$ has either {\it even} or {\it odd} parity. From
the hermiticity of $H_0$ the eigenvalues of the zeroth-order
Hamiltonian are real. Therefore, the n-th order energy correction
term, $\{E_k^{(n)}\}$, is also real. As can be verified by
substitution, the definite parity of the  $\psi_q^{(0)}(x)$
together with the antisymmetric parity of the perturbing
potential, i.e., $V(x)=-V(-x)$, generate a definite parity n-th
correction to the wavefunction. That is,
\beq
\psi_j^{(n)}(x)=\left(-1\right)^{j+n-1}\psi_j^{(n)}(-x),
\eeq
for: $j=1,2,\ldots$ and $n=0,1,\ldots$. One can now immediately
conclude that,
\begin{equation}
E_j^{(2n+1)}=0\,. \label{odd-E}
\end{equation}
Therefore,
\begin{equation}
E_j(|\lambda| < |\lambda_{bp}|)=\sum_{n=0}^\infty \lambda^{2n}
E_j^{(2n)}.
\end{equation}
Consequently for $\lambda=i|\lambda|$ the series
\begin{equation}
E_j(|\lambda| < |\lambda_{bp}|)=\sum_{n=0}^\infty (-1)^n|\lambda|^{2n} E_j^{(2n)}
\end{equation}
converges to real values.

The above analysis shows that the transition between exact and
spontaneously broken \PT - symmetry occurs at a branch point where
the non-hermiticity parameter ($|\lambda|$ in our case) reaches
the radius of convergence of the Rayleigh-Schr\"{o}dinger
perturbation expansion. As we will show below this phenomenon can
be observed experimentally.

The measurement of the radius of convergence of a
Rayleigh-Schr\"{o}dinger series expansion will be demonstrated in
a \PT - symmetric waveguide configuration. A \PT - symmetric
waveguide can be easily realized with a symmetric index guiding
profile and an antisymmetric gain/loss profile, i.e.,
$n(x)=n^*(-x)$ \cite{Christo}. We consider two coupled planar
waveguides depicted in Fig. \ref{fig1} for which the refractive
index varies only in the x direction. The direction of propagation
in the waveguides is taken to be the z axis. The wave equation for
the transverse-electric (TE) modes then reads:
\begin{figure}[htbp]
\includegraphics[width=3.5in,angle=0,scale=1,draft=false]{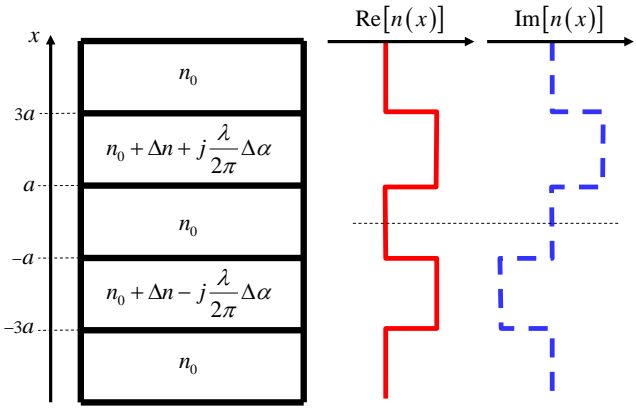}
\caption{(color online) A \PT-symmetric directional coupler. The
structure consists of two coupled slab waveguides. The real and
imaginary part of the refractive index is also portrayed. The
refractive index only varies in the x direction.}
\label{fig1}
\end{figure}
\beq
\label{waveEQ}
\left(\frac{\partial^2}{\partial
x^2}+k^2n(x)^2\right)\mathcal{E}_y(x)=\beta^2\mathcal{E}_y(x),
\eeq
where the y component of the electric field is given by:
$E_y(x,z,t)=\mathcal{E}_y(x)e^{i(\omega t-\beta z)}$,
$k=2\pi/\lambda$, and $\lambda$ is the vacuum wavelength. Clearly,
the wave equation for the y component of the electric field, i.e.,
Eq. (\ref{waveEQ}), is analogous to the one-dimensional
Schr\"{o}dinger equation:
$\left(-\frac{1}{2}\frac{\partial^2}{\partial
x^2}+V(x)\right)\Psi(x)=E\Psi(x)$, identifying the potential as
$V(x)=-\frac{1}{2}k^2n^2$ and the energy as:
$E=-\frac{1}{2}\beta^2$. As shown in Fig. \ref{fig1} we couple
between one gain-guiding waveguide (positive imaginary part of the
refractive index)  and one loss-guiding waveguide (negative
imaginary part of the refractive index) \cite{Siegman} in order to
create the \PT - symmetric structure. For simplicity we take the
separation between the two coupled waveguides to be the same as
the waveguides' width, i.e., $2a$. The effect of changing the
separation of two coupled \PT - symmetric waveguides was studied
in \cite{Christo}, and showed that as in regular (non \PT -
symmetric) directional couplers the coupling length is an
increasing monotonic function of the separation. The beat time
period usually used to describe quantum beating systems , e.g.,
\cite{Lovering} is exchanged in the study of waveguides by a beat
length: $L=2\pi/\Delta\beta$, where $\Delta\beta$ is the
difference between the propagation constants of the two modes. As
described below a different method of controlling the beat length
is by changing the non-hermiticity of the potential which can be
controlled by the gain(loss) coefficient: $\Delta\alpha$.

\begin{figure}[htbp]
\includegraphics[width=3.5in,angle=0,scale=1,draft=false]{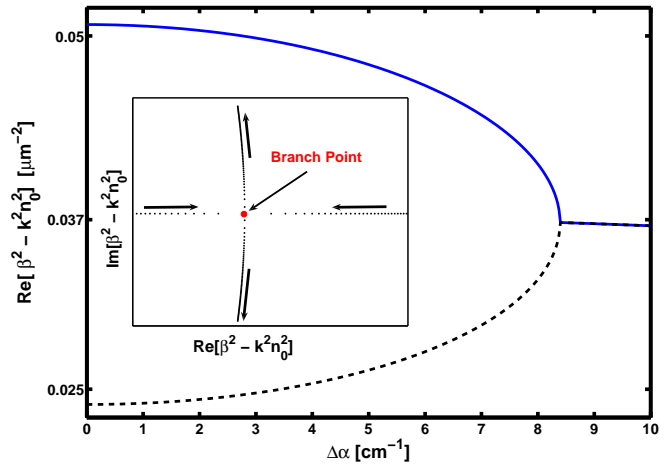}
\caption{(color online) The two trapped modes of the wave-guide
depicted in Fig. \ref{fig1} as a function of the non-hermiticity
parameter strength. The eigenmodes approach each other on the real
axis as $\Delta\alpha$ increases until a critical value of
$\Delta\alpha_c\sim8.4$ is reached. At the critical value one
finds a branch (exceptional) point where the two modes coalesce.
Beyond the branch point the directional coupler sustains one gain
guiding mode and one loss guiding mode.}
\label{fig2}
\end{figure}

In order to illustrate the control of the beat length using the
non-hermiticity parameter we choose the following parameters for
the waveguide structure shown in Fig. \ref{fig1}: The background
index is taken to be: $n_0=3.3$, the vacuum wavelength:
$\lambda=1.55\mu m$, the real index difference between the
waveguides and the background material: $\Delta n=10^{-3}$, and
the separation between the waveguides which equals the waveguides'
width: $2a=5\mu m$. The parameters are chosen such that each
waveguide contains only a single guided mode before we couple
them. The coupled guided modes are calculated by diagonalizing the
matrix representation of Eq. \ref{waveEQ} in a sine basis. The
"Hamiltonian" matrix is non-hermitian and one needs to take care
when normalizing the eigenvectors. We choose to normalize our
eigenvectors according to the so-called \textit{c-product}
\cite{nimrodrev}, i.e.,
$\left(\mathcal{E}_{n}|\mathcal{E}_{m}\right)=\langle
\mathcal{E}_{n}^*|\mathcal{E}_{m}\rangle=\delta_{n,m}$.

The coupled waveguides support two guided modes. The propagation
constants of the two modes are plotted in Fig. \ref{fig2} as a
function of the non-hermiticity parameter. Increasing
$\Delta\alpha$ causes the propagation constants of the two modes
to move towards each other up to a critical point, i.e., the
branch point. Beyond the branch point the propagation constants
become complex conjugates of one another. As long as \PT -
symmetry remains exact, i.e.,
$\Delta\alpha<\Delta\alpha_c\simeq8.4$, the two guided modes can
be classified according to the parity of the real part of the
transverse electric field just as in the non \PT - symmetric case.
The critical value of $\Delta\alpha$ corresponds to a branch
point, i.e., an exceptional point, where the two modes coalesce.
At the branch point both the propagation constants and the
corresponding electric field become equal. One can, therefore,
study the exceptional point in a \PT - symmetric waveguide by
varying only a \textit{single parameter}. Past the critical value
of $\Delta\alpha$, the waveguides support one gain-guiding mode
and one loss-guiding mode. The transverse field past the branch
point no longer retains the symmetry properties of the \PT
operator, but rather each of the two modes becomes localized in
one of the waveguides.

\begin{figure}[htbp]
\includegraphics[width=3.5in,angle=0,scale=1,draft=false]{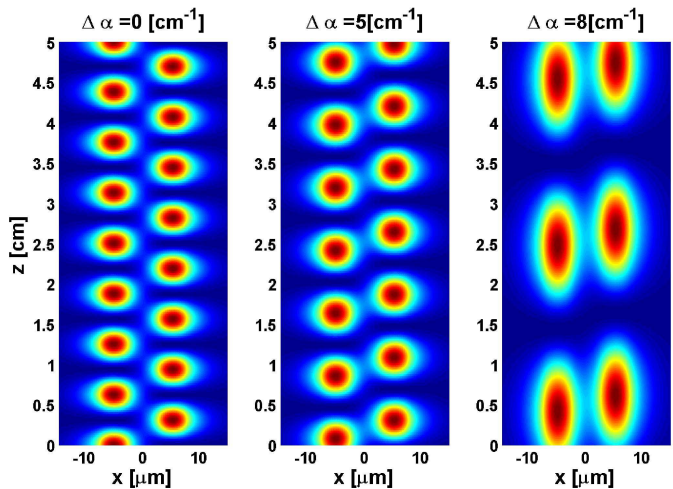}
\caption{(color online) The power distribution for a propagating
sum field consisting of the two guided modes, see Eq.
\ref{sumfield}. The propagating light is shown for three values of
$\Delta\alpha$. As can be readily observed, the beat length
(analogous to the beat time period in quantum mechanics) increases
as the value of $\Delta\alpha$ approaches the critical value. See
the text for further details.}
\label{fig3}
\end{figure}

The progression of the propagation constants on the real axis
towards the branch point can be visualized experimentally by
observing the beat length (time, in the corresponding quantum
mechanical problem) of the sum field for the \PT symmetric
waveguide. Fig. \ref{fig3} displays the power distribution,
\beq
\label{sumfield}
|E_y(x,z)|^2=\left|\frac{1}{\sqrt{2}}\left(\mathcal{E}_{1}(x)e^{-i\beta_1z}+\mathcal{E}_{2}(x)e^{-i\beta_2z}\right)\right|^2,
\eeq
for three values of $\Delta\alpha$. As the value of $\Delta\alpha$
approaches the critical value the beat length increases. This is a
direct observation of the movement of the propagation constants
towards each other on the real axis. Near the critical value of
$\Delta\alpha$, i.e., the exceptional point, the sum field no
longer oscillates between the waveguides but rather travels in
both waveguides simultaneously. Therefore, by tuning the value of
$\Delta\alpha$ one can visualize the movement of the eigenmodes
towards the branch point or away from it, as seen in Fig.
\ref{fig3}.

The effect of the non-hermiticity on the beat length is clearly
reflected in Fig. \ref{fig3}. However, this visualization of the
branch point is not the only information one can get from such an
experiment. By finding what is the critical value of
$\Delta\alpha$ which corresponds to the branch point one can infer
the radius of convergence for any symmetric waveguide structure
with an antisymmetric perturbation. Therefore, while the
visualization is done on a \PT symmetric waveguide, the critical
value of $\Delta\alpha$ corresponds to the maximum value of an
added antisymmetric index profile which can still be treated
within perturbation theory.

In conclusion, a proof of the reality of the spectrum of \PT -
symmetric Hamiltonians has been given using the
Rayleigh-Schrodinger perturbation theory. We showed that the
transition between exact and spontaneously broken \PT - symmetry
occurs at a branch point and the transition can be visualized in a
\PT symmetric waveguide structure. Such a system has many
advantages for the study of exceptional points as the movement
towards the branch point is controlled by a single parameter. We
also show that by such a measurement one can find the radius of
convergence for the family of waveguides where a symmetric index
profile is supplemented with an antisymmetric index profile. This
antisymmetric profile need not be imaginary. The knowledge of the
radius of convergence is important if one wishes to treat the
added antisymmetric index as a perturbation.

NM acknowledges the financial support of Israel Science Foundation
(grant No. 890015). UG has been supported by the German Research
Foundation DFG, grant GE 682/12-3. SK acknowledges the support of
the ministry of science, culture, and sports.

\bibliographystyle{unsrt}

\end{document}